\documentclass{Interspeech}

\interspeechcameraready

\title{PromptEVC: Controllable Emotional Voice Conversion with\\ Natural Language Prompts}

\author[affiliation={1,2}]{Tianhua}{Qi}
\author[affiliation={1,2}]{Shiyan}{Wang}
\author[affiliation={1,2}]{Cheng}{Lu}
\author[affiliation={3}]{Tengfei}{Song}
\author[affiliation={3}]{Hao}{Yang}
\author[affiliation={3}]{Zhanglin}{Wu}
\author[affiliation={1,2}]{\\Wenming}{Zheng}
\affiliation{Key Laboratory of Child Development and Learning Science (Southeast University)}{\\Ministry of Education}{Nanjing 210096, China}
\affiliation{School of Biological Science and Medical Engineering}{Southeast University}{China}
\affiliation{}{Huawei Translation Service Center}{China}
\email{qitianhua@seu.edu.cn, yanghao30@huawei.com, wenming\_zheng@seu.edu.cn}

\usepackage{comment}
\usepackage{multirow}

\begin{document}

\maketitle

\begin{abstract}
Controllable emotional voice conversion (EVC) aims to manipulate emotional expressions to increase the diversity of synthesized speech. Existing methods typically rely on predefined labels, reference audios, or prespecified factor values, often overlooking individual differences in emotion perception and expression. In this paper, we introduce PromptEVC that utilizes natural language prompts for precise and flexible emotion control. To bridge text descriptions with emotional speech, we propose emotion descriptor and prompt mapper to generate fine-grained emotion embeddings, trained jointly with reference embeddings. To enhance naturalness, we present a prosody modeling and control pipeline that adjusts the rhythm based on linguistic content and emotional cues. Additionally, a speaker encoder is incorporated to preserve identity. Experimental results demonstrate that PromptEVC outperforms state-of-the-art controllable EVC methods in emotion conversion, intensity control, mixed emotion synthesis, and prosody manipulation.
Speech samples are available at \url{https://jeremychee4.github.io/PromptEVC/}.
\end{abstract}

\section{Introduction}
Emotional voice conversion (EVC) aims to modify the emotional tone of spoken utterances while preserving their linguistic content and speaker identity~\cite{zhou2022emotional}.
This technique facilitates emotional communication, enhances user experiences in human-machine interactions, and contributes to more immersive virtual environments~\cite{triantafyllopoulos2023overview}.
Conventional EVC systems primarily rely on autoencoders~\cite{chen2022speaker, zhu2023emotional}, achieving significant improvements in speech quality~\cite{kreuk2022textless, qi2024pavits}. However, their limited variability in synthesized voices restricts the diversity of emotional expressions~\cite{qi2024towards}.
To overcome this limitation, controllable EVC has attracted significant attention by enabling the manipulation of emotional expression and addressing the one-to-many problem via reference speech or intuitive values.

Reference audio-based controllable EVC methods~\cite{kim2020emotional,zhou2021seen,lei2022msemotts}, such as ZEST~\cite{dutta2024zero}, extract emotional features from reference speech through a reference encoder, transferring these features to the source audio to enable emotional transformations without predefined labels or other form of information. However, these approaches often require manual selection of reference speech, complicating their practical application.
In contrast, intuitive values-based controllable EVC methods~\cite{qi2024towards, xue2018voice} adjust vocal expressions based on target emotion labels and numerical factors, to simplify operation. For instance, Emovox~\cite{zhou2022emotion} employs the relative attribute ranking (RAR)~\cite{parikh2011relative} to measure intra-class and inter-class distances of acoustic features across various emotions, offering finer control over emotional intensity. Mixed-EVC~\cite{zhou24_odyssey} evaluates differences between speech samples of different emotions, allowing model to generate mixtures of emotions (e.g., ``\textit{happy with surprise}") by manually defining an attribute vector. However, since human subjectively interpret emotions in different ways~\cite{mower2009interpreting}, relying solely on numerical values may not align with individual expectations, potentially affecting controllability.

Text descriptions, as a direct form of human communication~\cite{de1981introduction}, can be used to convey multiple dimensions of speech emotion~\cite{juslin2005vocal}, such as \textit{rhythm}, \textit{volume}, \textit{emotional intensity}, and \textit{mixed emotions}. By using text descriptions as prompts, vocal expressions can be flexibly edited, thus circumventing the limitations of previous methods that rely on predefined labels, reference speeches, or numeric values.
To handle similar issue, several solutions have been explored in text-to-speech (TTS)~\cite{guo2023prompttts,lengprompttts} and voice conversion (VC)~\cite{yao2024promptvc}. However, this controllable approach has not yet been extended to emotion conversion.

In this paper, we propose PromptEVC that leverages natural language prompts to control vocal expressions during emotion conversion, thereby enhancing both the controllability and emotional diversity of synthesized speech.
To establish a mapping between textual descriptions of target emotions and their corresponding speech, we introduce an emotion descriptor that generates an initial emotion embedding from the provided description. 
Since speech emotion is inherently suprasegmental~\cite{govind2011neutral}, we further introduce a prompt mapper to refine the emotion embedding's granularity, which is trained jointly with reference embeddings extracted from reference speech.
To improve the naturalness of generated speech, we introduce a prosody modeling and control pipeline. Within this pipeline, the duration regulator adjusts the rhythm based on linguistic tokens and latent emotional information, while the prosody predictor synthesizes natural prosody by integrating multiple dimensions of emotional speech with linguistic content.
Additionally, a speaker encoder with an F0 constraint is utilized to preserve speaker identity, particularly when manipulating intonation.
This system allows users to describe their desired emotional expression in natural language, such as ``\textit{Please convert it into a very happy tone with a slightly faster tempo, and a touch of surprise would be great}'', which enables precise control over emotions while effectively mitigating individual differences.
\begin{figure*}[htb]
  \centering
  \includegraphics[width=2\columnwidth]{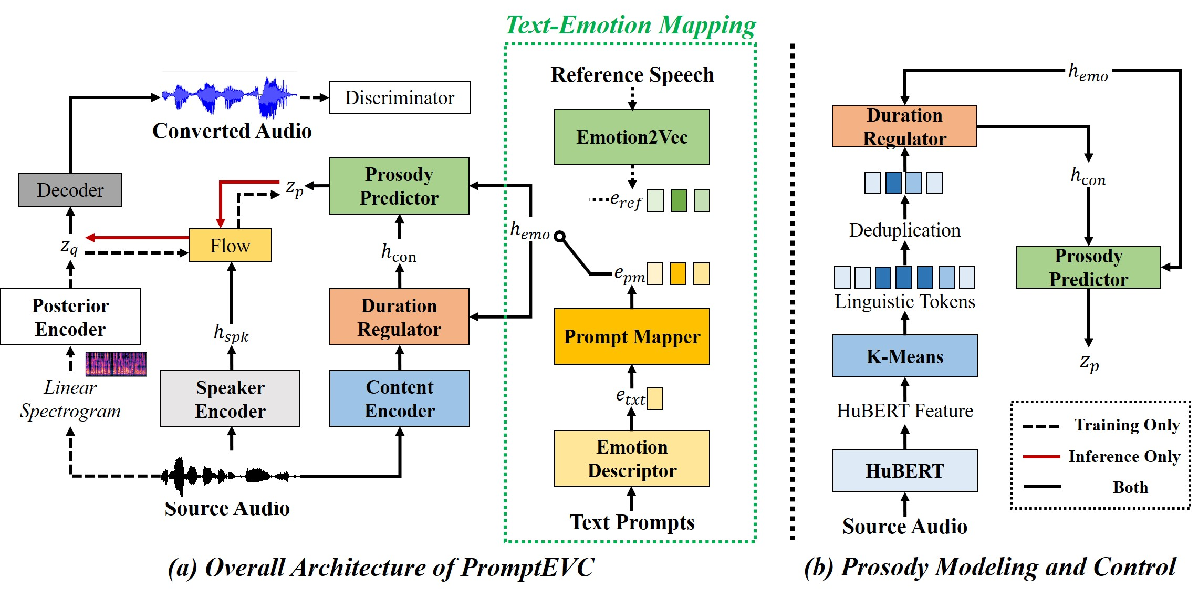}
  \caption{Diagram of proposed PromptEVC, depicting the overall architecture (a) and the prosody modeling and control module (b).}
  \label{fig:training}
\end{figure*}

\section{Proposed Method}
As illustrated in Figure 1(a), PromptEVC is built upon conditional variational autoencoder (CVAE), consisting of: 1) a text-emotion mapping module, 2) a prosody modeling and control module, 3) a speaker encoder for identity preservation, and 4) a decoder for waveform reconstruction.

\subsection{Text-emotion mapping}
Although natural language prompts offer considerable variability over previous approaches, there remains an information mismatch compared to original speech signals, which convey emotions through suprasegmental features such as intonation, loudness, and rhythm—elements not captured by the verbal modality. To bridge this gap and create a more effective mapping between these two modalities, we propose the emotion descriptor. This component processes the user's textual description to generate an initial coarse-level emotion embedding, denoted as $e_{txt}$. We then introduce a prompt mapper, which jointly models $e_{txt}$ with an reference embedding, $e_{ref}$, extracted from reference speech using Emotion2Vec\footnote{\url{https://github.com/ddlBoJack/emotion2vec}}~\cite{ma-etal-2024-emotion2vec}, a large-scale self-supervised pre-trained model for speech emotion representation. This approach allows us to derive a fine-grained emotion embedding, $e_{pm}$, which better aligns with the textual description. Both $e_{pm}$ and $e_{ref}$ serve as intermediate emotional representations, denoted as $h_{emo}$, within the system, with $e_{pm}$ being used during inference.

\textit{Emotion descriptor}: 
Natural language prompts can vary significantly in length and content, with users potentially prioritizing different aspects of emotion, such as emotional category, intensity, mixed emotions, and prosodic variations. To address this, we construct an emotion descriptor using RoBERTa~\cite{Liu2019RoBERTaAR}, an advanced BERT framework, to generate a fixed-length initial emotion embedding, $e_{txt}$. It involves a pre-trained RoBERTa model\footnote{\url{https://huggingface.co/FacebookAI/roberta-base}} followed by a linear projection layer.

\textit{Prompt mapper}:
To better align $e_{ref}$ extracted from emotional speech and capture diverse variations in human speech, we employ a diffusion model to predict the fine-grained emotion representation $e_{pm}$ conditioned on $e_{txt}$. In the diffusion process, the reference representation $\mathbf{x}_0$ ($e_{ref}$) is progressively transformed into Gaussian noise under the noise schedule $\mathcal \beta$ at each time step $t$, as described in Eq. (1).
\begin{equation}
\mathrm{d} \mathbf{x}_t=-\frac{1}{2} \beta_t \mathbf{x}_t \mathrm{~d} t+\sqrt{\beta_t} \mathrm{~d} w_t, \quad t \in[0,1]
\end{equation}
where ${~d} w_t$ denotes the increment of a standard Wiener process, which introduces randomness into the evolution of $\mathbf{x}_t$.

For the denoising process, the prompt mapper is trained to estimate the gradients of the log-density of noisy data $\nabla \log p_t\left(\mathbf{x}_t\right)$ by predicting $\mathbf{x}_0$, as shown in Eq. (2).
\begin{equation}
\mathrm{d} \mathbf{x}_t=-\frac{1}{2}\left(\mathbf{x}_t+\nabla \log p_t\left(\mathbf{x}_t\right)\right) \beta_t \mathrm{~d} t, \quad t \in[0,1]
\end{equation}
This process is implemented using multiple Transformer Encoder layers, and the training loss is defined in Eq. (3).
\begin{equation}
    \mathcal{L}_{pm}=\mathbb{E}_{\mathbf{x}_0, \epsilon \sim \mathcal{N}(0, I), t} \left[ ||\boldsymbol{\epsilon}-\boldsymbol{\epsilon}_\theta (\mathbf{x}_t, e_{txt}, t)|| \right]^2
\end{equation}
where $\epsilon$ represents the diffusion noise sampled from a standard normal distribution, $\theta$ represents a set of trainable parameters. 

\subsection{Prosody modeling and control}
Controllable EVC systems often require modifying acoustic details that may impact the naturalness of generated speech. To address this, we propose a prosody modeling and control pipeline, as illustrated in Figure 1(b), to ensure the quality of prosodic variations during emotion conversion.

\textit{Linguistic tokens}:
The pre-trained HuBERT model\footnote{\url{https://huggingface.co/facebook/hubert-base-ls960}}~\cite{hsu2021hubert} is employed to extract continuous hidden representations from emotional speech, capturing rich information such as content, prosody, and timbre ~\cite{lakhotia2021generative}. To further distill the linguistic content, we apply K-means clustering to quantize these representations into discrete linguistic tokens.

\textit{Duration regulator}:
The linguistic tokens derived from HuBERT with K-means clustering often exhibit contiguous repetitions, reflecting phoneme articulation duration and its impact on speech emotion. To enhance the rhythmic modeling of emotional speech, these tokens are deduplicated and fed into the duration regulator to generate a duration-aware content representation, denoted as $h_{con}$. It is composed of six 1-D CNN layers followed by a linear projection layer.
\begin{equation}
    \mathcal{L}_{rhy}=\frac{1}{n} \sum_{i=1}^N \log \cosh \left(\hat{y}_i-\log \left(y_i+1\right)\right)
\end{equation}
where $y_i$ denotes the ground truth duration of the $i$-th sample, and $\hat{y_i}$ represents its predicted value.

\textit{Prosody predictor}: 
Prosody, as a fundamental element of naturalness and emotional expressiveness, includes acoustic attributes such as pitch, duration, and energy~\cite{mozziconacci2002prosody}. However, due to its suprasegmental nature, relying solely on text- or phoneme-level features is insufficient for modeling. To capture prosodic patterns with high temporal resolution, the prosody predictor learns a joint representation of content ($h_{con}$) and emotion ($h_{emo}$). It is achieved through two 1-D CNN layers followed by a linear projection layer.

\subsection{Speaker identity preservation}
Controllable EVC systems often prioritize generating speech that aligns with the target emotion distribution, potentially neglecting the speaker's identity. Recognizing the pivotal role of fundamental frequency ($F_0$) in speaker modeling~\cite{wolf1972efficient}, we enhance a pre-trained speaker verification model\footnote{\url{https://huggingface.co/microsoft/wavlm-base-sv}} by incorporating two 1-D CNN layers and a linear projection layer with an $F_0$ constraint.
\begin{equation}
    \mathcal{L}_{spk}=||\log F_0-\log \hat{F_0}||_2
\end{equation}
where $F_0$ represents the ground truth fundamental frequency, $\hat{F_0}$ denotes the predicted value.

\subsection{Waveform reconstruction}
To enhance the vocal expressiveness of PromptEVC, we introduce a posterior encoder during training phase, which learns spectral details from the linear spectrogram and provides the complex posterior distribution $z_q$ for CVAE. A reversible normalizing flow, conditioned on $h_{spk}$, is used to capture complex speech emotions. The decoder then generates waveforms from $z_q$ and employs adversarial learning to improve the naturalness of content and emotion.

\subsection{Final loss}
By combining CVAE with adversarial training, the overall loss function is formulated as:
\begin{equation}
    \mathcal{L}=\mathcal{L}_{sp}+\mathcal{L}_{f}+\mathcal{L}_{adv}(G)+\mathcal{L}_{pm}+\mathcal{L}_{rhy}+\mathcal{L}_{spk}
\end{equation}
\begin{equation}
    \mathcal{L}(D)=\mathcal{L}_{adv}(D)
\end{equation}
where $\mathcal{L}_{sp}$ minimizes the $\mathcal{L}_1$ distance between the synthesized and target spectrograms, $\mathcal{L}_{f}$ minimizes the $\mathcal{L}_1$ distance between feature maps extracted from intermediate layers of each discriminator, $\mathcal{L}_{adv}(G)$ and $\mathcal{L}_{adv}(D)$ represent the adversarial loss for the Generator and Discriminator, respectively. 

\section{Experiments}
\subsection{Experimental setup}
\textbf{Datasets.} We conduct emotion conversion using the TextrolSpeech dataset~\cite{ji2024textrolspeech}, which consists of 330 hours of audio, 236,203 pairs of natural text descriptions with corresponding speech samples. The dataset includes 42,909 emotional speech samples, categorized as follows: angry (18.75\%), contempt (6.25\%), disgusted (12.5\%), fear (6.25\%), happy (18.75\%), sad (18.75\%), and surprised (18.75\%). The remaining non-emotional samples are labeled as neutral. Each text description comprises five style factors: gender ($G$), pitch ($P$), speaking speed ($S$), volume ($V$), and emotion category ($E_{cg}$).\\
\textbf{Data preparation.} 
Following the approach in~\cite{zhou2022emotion}, we measure emotional intensity ($ E_{in}$) using  RAR\cite{parikh2011relative}. After sorting the values in descending order, we categorize the top 30\%, middle 40\%, and bottom 30\% as high, medium, and low emotional intensity levels, respectively. We then modify the style prompts from TextrolSpeech using ChatGPT-4 to better align with the emotion conversion task. Specifically, we input the original descriptions along with each factors (emotional category, emotional intensity, pitch, speaking speed, and volume) and their corresponding categories into ChatGPT-4, instructing it to rewrite the natural language prompts while removing any gender-specific information, as preserving speaker identity is crucial in EVC. Additionally, we enhance the complexity of descriptions by leveraging ChatGPT-4 to grammatically restructure them.\\
\textbf{Implementation details.} 
The proposed architecture builds upon VITS ~\cite{kim2021conditional}, employing the AdamW optimizer with an initial learning rate of 2e-4, a dropout probability of 0.1, and a learning rate decay factor of 0.999875. The dictionary size of the extracted HuBERT units is set to 100. The model is pre-trained for 500 epochs on the neutral dataset and fine-tuned for an additional 200 epochs on the emotional dataset.\\
\textbf{Models for comparison.}
We train the following models to evaluate the performance of the proposed method.
\begin{itemize}
    \item Textless-EVC (baseline)~\cite{kreuk2022textless}: A predefined label-based EVC model that uses HuBERT units for emotion conversion.
    \item ZEST (baseline)~\cite{dutta2024zero}: A reference audio-based controllable EVC model that utilizes self-supervised techniques to predict pitch contours from source and reference speech.
    \item Emovox (baseline)~\cite{zhou2022emotion}: An intuitive values-based controllable EVC model that enables emotional intensity control through the target emotion label and intensity value.
    \item Mixed-EVC (baseline)~\cite{zhou24_odyssey}: An intuitive values-based controllable EVC model that synthesizes mixed emotional expressions via the manually specified attribute vector.
    \item PromptEVC (proposed): The proposed model that leverages natural language prompts to encompass multiple emotional aspects, allowing flexible control over emotion conversion.
\end{itemize}
\begin{table*}[th]
  \caption{Quantitative comparison of converted speech with previous methods in terms of objective and subjective evaluations.}
  \label{tab:comparison}
  \centering
    \begin{tabular}{cccccccc}
    \hline
    \toprule
   \multirow{2}{*}{Model} & \multirow{2}{*}{Setting}          & \multicolumn{4}{c}{Objective Evaluation} & \multicolumn{2}{c}{Subjective Evaluation} \\ \cline{3-8} 
                       &                                   & MCD $\downarrow$    & CER $\downarrow$    & $\text{RMSE}_{F_0}$ $\downarrow$   & $\text{ACC}_{cls}$ $\downarrow$  & Naturalness $\uparrow$          & Similarity $\uparrow$         \\ \hline
Textless-EVC & Predefined Labels                 & 9.00 & 5.73 & 64.82 & 98.66 & 3.67±0.11 & 74.10\% \\
ZEST         & Reference Speech                  & 8.45 & 5.32 & 45.59 & 98.72 & 3.58±0.13 & 75.67\% \\
Emovox       & \multirow{2}{*}{Intuitive Values} & 5.26 & 5.07 & 46.69 & 98.67 & 3.95±0.14 & 76.50\% \\
Mixed-EVC    &                                   & 6.32 & 5.21 & 48.22 & 98.63 & 3.89±0.19 & 76.07\% \\ \hline
PromptEVC (proposed) &
  \multirow{4}{*}{Text Description} &
  \textbf{4.70} &
  \textbf{4.09} &
  \textbf{42.58} &
  \textbf{99.32} &
  \textbf{4.22±0.08} &
  \textbf{81.30\%} \\
\multicolumn{1}{r}{w/o Prompt Mapper}        &                                   & -    & 4.24    & 43.87     & 98.58       & 3.83±0.19         & -       \\
\multicolumn{1}{r}{w/o Prosody Predictor}       &                                   & -    & 4.53    & 44.54     & 99.05      & 4.02±0.11        & -      \\
\multicolumn{1}{r}{w/o Speaker Encoder}       &                                   & -   & 4.12   & 49.36    & 99.16      & 4.16±0.09        & -          \\ 
   \bottomrule
   \hline
    \end{tabular}
\end{table*}
\begin{table}[th]
\caption{The classification accuracy (\%) across multiple attributes, evaluated by pre-trained classifier.}
  \label{tab:accuracy}
  \centering
 \begin{tabular}{ccccccc}
\hline
 \toprule
 & $E_{in}$ & $E_{mx}$ & $P$ & $S$ & $V$ \\ \hline
                 & 77.58  & 61.25  & 86.41 & 89.72 & 91.52 \\
             \bottomrule   
               \hline

\end{tabular}
\end{table}
\begin{table}[th]
\caption{The average classification accuracy (\%) across multiple attributes, evaluated by five participants.}
  \label{tab:accuracy}
  \centering
 \begin{tabular}{ccccccc}
\hline
 \toprule
 & $E_{in}$ & $E_{mx}$ & $P$ & $S$ & $V$ \\ \hline
Proposed       & 72.90  & 60.04  & 84.38 & 89.13 & 89.98 \\
Ground Truth    & 78.07  & -  & 88.87 & 90.64 & 92.45 \\ 
 \bottomrule
\hline
\end{tabular}
\end{table}
\subsection{Model performance}
As shown in Table 1, we evaluate the models using several metrics, including Mel cepstral distortion (MCD), character error rate (CER, \%), root mean squared error of log $F_0$ ($\text{RMSE}_{F_0}$), and classification accuracy from a pre-trained speech emotion recognition (SER) model~\cite{gao23g_interspeech} ($\text{ACC}_{cls}$, \%) for objective evaluation. For subjective evaluation, we conduct a mean opinion score (MOS) test with 25 participants, each evaluating 132 utterances in total, to assess the naturalness and emotional similarity of converted audios.

The results show that proposed PromptEVC performs competitively in both objective and subjective evaluations. In terms of MCD, $\text{RMSE}_{F_0}$, and MOS, both Emovox and PromptEVC outperform other controllable EVC models. This is due to their ability to capture and generate prosodic variations at a fine-grained level, underscoring the importance of detailed emotional descriptions in emotion conversion. However, Mixed-EVC performs suboptimally, which may be attributed to the challenges posed by mixed emotions, commonly present in social interactions but difficult to model explicitly. Additionally, the CER of PromptEVC is reduced by approximately 1\% compared to baseline methods, indicating that our proposed PromptEVC can effectively mitigate mispronunciations and skipping-words while controlling emotional transitions.

\subsection{Ablation study}
We perform an ablation study to assess the impact of different components. Specifically, we sequentially remove the prompt mapper, prosody predictor, and speaker encoder, evaluating performance using CER, $\text{RMSE}_{F_0}$, and $\text{ACC}_{cls}$ for objective evaluation, along with MOS for subjective evaluation.
As shown in Table 1, all performance degrades when different components are removed. Removing the prompt mapper increases $\text{RMSE}_{F_0}$ by 1.29 and decreases $\text{ACC}_{cls}$ and naturalness by 0.74\% and 0.39, respectively. This suggests that directly predicting emotion embeddings from text is insufficient and leads to unnatural prosody. Replacing the prosody predictor with concatenation results in a 0.44\% increase in CER and a 0.2 decrease in naturalness, as this substitution lacks in integrating linguistic content, speech rhythm, and emotional cues. Finally, removing the speaker encoder causes a significant rise in $\text{RMSE}_{F_0}$, indicating that the loss of the specific constraint ($\mathcal{L}_{spk}$) hinders the system's capacity to preserve  identical characteristics.
\begin{figure}[htb]
  \centering
  \includegraphics[width=\linewidth]{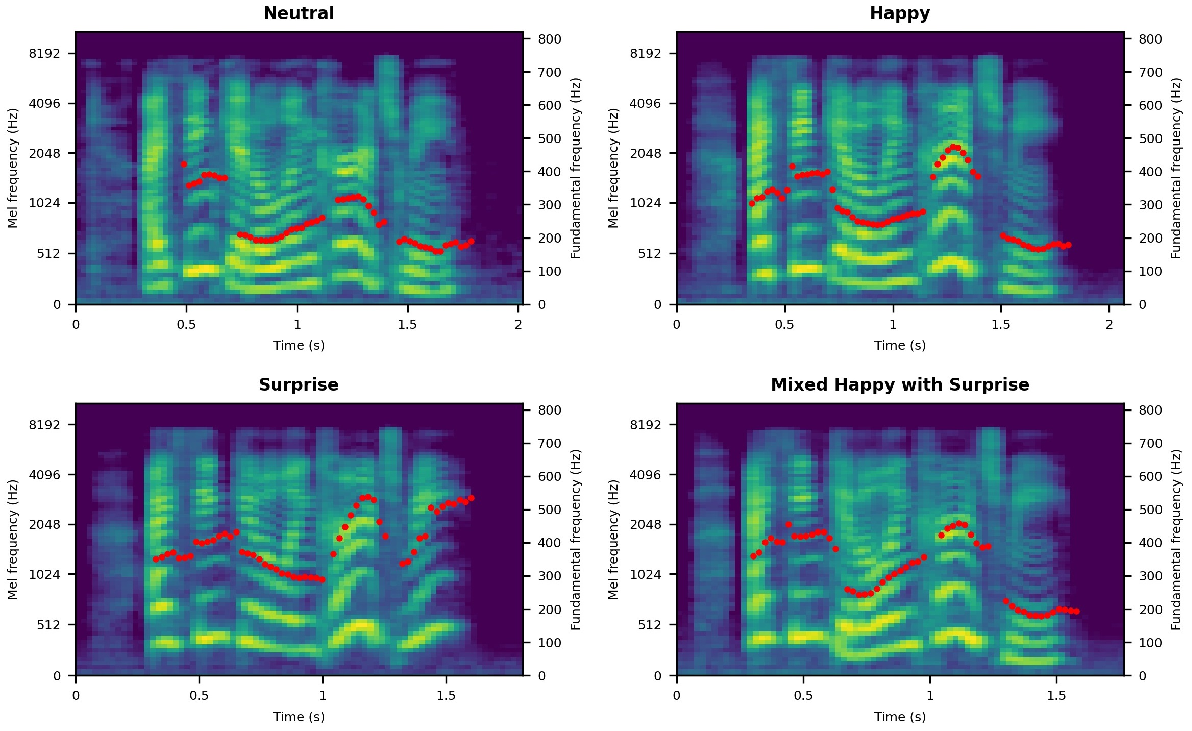}
  \caption{Mel-spectrograms and pitch contours of the converted audios for different natural language prompts: ``Convert this neutral utterance into a happy/surprise tone'' and ``Inject some happiness into this sentence with a hint of surprise.''}
  \label{fig:mel}
\end{figure}
\subsection{Controllability of natural language prompts}
To demonstrate the controllability of natural language prompts in emotion conversion, we conduct both objective and subjective evaluations across multiple attributes, including emotional intensity, mixed emotions ($E_{mx}$), pitch, speech speed, and volume. For objective evaluation, we pre-train a classifier to categorize each factor and compute the accuracy, as shown in Table 2. For subjective evaluation, five participants annotate the test audios alongside the ground truth, and the average classification accuracy is calculated, as shown in Table 3. The results indicate that the emotional speech generated using natural language prompts closely aligns with human expectations. Furthermore, Figure 2 presents synthesized Mel spectrograms with pitch contours, illustrating how altering the textual description influences the output. This suggests that PromptEVC can adaptively adjust prosodic variations to meet diverse individual demands.

\section{Conclusion}
In this paper, we propose PromptEVC to control subtle emotional expressions in emotional voice conversion (EVC) through natural language prompts. 
By incorporating emotion descriptor, prompt mapper with prosody modeling and control module, PromptEVC effectively learns fine-grained emotional representations across multiple attributes.
Experimental results on TextrolSpeech demonstrate that PromptEVC significantly improves both controllability and diversity. In the future, we will explore streaming EVC to enable real-time applications.

\section{Acknowledgment}
This work was supported in part by the National Key R\&D Project under the Grant 2022YFC2405600, 
in part by the NSFC under the Grant U2003207, 
in part by the YESS Program by JSAST under the Grant JSTJ-2023-XH033, 
in part by the China Postdoctoral Science Foundation under the Grant 2023M740600, 
in part by the Jiangsu Province Excellent Postdoctoral Program, 
and in part by the Postgraduate Research \& Practice Innovation Program of Jiangsu Province (KYCX25\_0515).

\bibliographystyle{IEEEtran}
\bibliography{mybib}

\end{document}